\documentclass[11pt]{IEEEtran}

\usepackage{cite,graphicx,psfrag,amssymb,amsmath,url,marvosym}
\newtheorem{theorem}{Theorem}
\newtheorem{lemma}{Lemma}
\newtheorem{corollary}{Corollary}

\def\lefthand{\reflectbox{\Pointinghand}}
\def\definedas{\triangleq}
\def\opt{{\mathop{\rm opt}}}
\def\root{{\mathop{\rm root}}}
\def\boldl{{\mbox{\boldmath $l$}}}
\def\boldsl{{\mbox{\scriptsize \boldmath $l$}}}
\def\p{{\mbox{\boldmath $p$}}}
\def\sp{{\mbox{\scriptsize \boldmath $p$}}}
\def\setA{{\mathcal A}}
\def\code{{\mbox{\boldmath $c$}}}

\def\Cost{L}
\def\E{{\mathbb E}}
\def\lu{\lambda}
\def\q{a}
\def\ll{(}
\def\lr{)}
\def\L{{{\mathcal L}_n}}
\def\setN{{\mathcal N}}
\def\P{{\mathbb P}}
\def\setP{{\mathcal P}}
\def\X{{\mathcal X}}
\def\Z{{\mathbb Z}}

\begin{document}
\bibliographystyle{IEEEtran} 
\title{Redundancy-Related Bounds for Generalized Huffman Codes}
\author{Michael~B.~Baer,~\IEEEmembership{Member,~IEEE}%
\thanks{Material in this paper was presented at the 2006
  International Symposium on Information Theory, Seattle, Washington,
  USA and the 2008 International Symposium on Information Theory, Toronto, Ontario, Canada.}%
\thanks{The author is with Vista Research, 4 Lower Ragsdale Drive, Suite 220, Monterey, California 93940  USA (email: calbear@ieee.org).}
\thanks{This work has been submitted to the IEEE for possible publication. Copyright may be transferred without notice, after which this version may no longer be accessible.}
}
\maketitle

\begin{abstract}
This paper presents new lower and upper bounds for the compression rate of binary prefix codes optimized over memoryless sources according to various nonlinear codeword length objectives.  Like the most well-known redundancy bounds for minimum average redundancy coding --- Huffman coding --- these are in terms of a form of entropy and/or the probability of an input symbol, often the most probable one.  The bounds here, some of which are tight, improve on known bounds of the form $\Cost \in [H,H+1)$, where $H$ is some form of entropy in bits (or, in the case of redundancy objectives,~$0$) and $\Cost$ is the length objective, also in bits.  The objectives explored here include exponential-average length, maximum pointwise redundancy, and exponential-average pointwise redundancy (also called $d$\textsuperscript{th} exponential redundancy).  The first of these relates to various problems involving queueing, uncertainty, and lossless communications; the second relates to problems involving Shannon coding and universal modeling.  For these two objectives we also explore the related problem of the necessary and sufficient conditions for the shortest codeword of a code being a specific length.exponential redundancy).  The first of these relates to various problems involving queueing, uncertainty, and lossless communications; the second relates to problems involving Shannon coding and universal modeling.  For these two objectives we also explore the related problem of the necessary and sufficient conditions for the shortest codeword of a code being a specific length.
\end{abstract}

\begin{keywords}
Huffman codes, optimal prefix code, queueing, R\'{e}nyi entropy, Shannon codes, worst case minimax redundancy.
\end{keywords} 

\section{Introduction} 

Since Shannon introduced information theory, we have had entropy bounds for the expected codeword length of optimal lossless fixed-to-variable-length binary codes.  The lower bound is entropy, while the upper bound is one bit greater --- corresponding to a maximum average redundancy of one bit for an optimal code --- thus yielding \textit{unit-sized} bounds.  The upper bound follows from the suboptimal Shannon code, a code for which the codeword length of an input of probability $p$ is $\lceil -\log_2 p \rceil$\cite{Shan}.

Huffman found a method of producing an optimal code by building a tree in which the two nodes with lowest weight (probability) are merged to produce a node with their combined weight summed\cite{Huff}.  On the occasion of the twenty-fifth anniversary of the Huffman algorithm, Gallager introduced bounds in terms of the most probable symbol which improved on the unit-sized redundancy bound\cite{Gall}.  Since then, improvements in both upper and lower bounds given this most probable symbol\cite{John,CGT,MoAb,CaDe1,Mans} have yielded bounds that are tight when this symbol's probability is at least $1/127$ (and close-to-tight when it has lower probability).  Tight bounds also exist for upper and lower bounds given the less-specific information of an \textit{arbitrary} symbol's probability\cite{YeYe2,MPK}.  Such bounds are useful for quickly bounding the performance of an optimal code without running the algorithm that would produce the code.  The bounds are for a fixed-sized input alphabet; \textit{asymptotic} treatment of redundancy for block codes of growing size, based on \textit{binary} memoryless sources, is given in \cite{Szpa}.

Others have given consideration to objectives other than expected codeword length\cite[\S 2.6]{Abr01}.  Many of these nonlinear objectives, which have a variety of applications, also have unit-sized bounds but have heretofore lacked tighter closed-form bounds achieved using a symbol probability and, if necessary, some form of entropy.  We address such problems here, finding upper and lower bounds for the optimal codes of given probability mass functions for nonlinear objectives.  ``Optimal'' in this paper refers to optimality over the objective in question, not necessarily over the linear objective of expectation.

A lossless binary prefix coding problem takes a probability mass function 
$\p = \{p_i\}$, defined for all $i$ in the input alphabet $\X$, and
finds a binary code for~$\X$.  Without loss of generality, we consider
an $n$-item source emitting symbols drawn from the alphabet $\X = \{1,
2, \ldots, n\}$ where $\{p_i\}$ is the sequence of probabilities for
possible symbols ($p_i > 0$ for $i \in \X$ and $\sum_{i \in \X} p_i
= 1$) in monotonically nonincreasing order ($p_i \geq p_j$ for
$i<j$).  Thus the most probable symbol is~$p_1$.  The source symbols are
coded into binary codewords.  The codeword $c_i \in \{0,1\}^*$ in
code~$\code$, corresponding to input symbol~$i$, has length $l_{i}$,
defining length vector~$\boldl$.

The goal of the traditional coding problem is to find a prefix
code minimizing expected codeword length $\sum_{i \in \X}
p_i l_{i}$, or, equivalently, minimizing average redundancy
\begin{equation}
\bar{R}(\boldl,\p) \definedas \sum_{i \in \X} p_i l_{i} - H(\p) =
\sum_{i \in \X} p_i (l_{i}+\lg p_i)
\label{minavg}
\end{equation}
where $H$ is $-\sum _{i \in \X} p_i\lg p_i$ (Shannon entropy) and $\lg
\definedas \log_2$.  A \textit{prefix code} --- also referred to as a
\textit{comma-free code}, a \textit{prefix-free code}, or an
\textit{instantaneous code} --- is a code for which no codeword begins
with a sequence that also comprises the whole of a second codeword.

This problem is equivalent to finding a minimum-weight external path
among all rooted binary trees, due to the fact that every prefix code
can be represented as a binary tree.  In this tree representation,
each edge from a parent node to a child node is labeled $0$ (left) or
$1$ (right), with at most one of each type of edge per parent node.  A
leaf is a node without children; this corresponds to a codeword, and
the codeword is determined by the path from the root to the leaf.
Thus, for example, a leaf that is the right-edge ($1$) child of a
left-edge ($0$) child of a left-edge ($0$) child of the root will
correspond to codeword~$001$.  Leaf depth (distance from the root) is
thus codeword length.  If we represent external path weight as
$\sum_{i \in \X} w(i) l_{i}$, the weights are the probabilities (i.e.,
$w(i) = p_i$), and, in fact, we refer to the problem inputs as
$\{w(i)\}$ for certain generalizations in which their sum, $\sum_{i
  \in \X} w(i)$, need not be~$1$.

If formulated in terms of $\boldl$, the constraints on the
minimization are the integer constraint (i.e., that codes must be of integer
length) and the Kraft inequality\cite{McMi}; that is, the set of
allowable codeword length vectors is
$$\L \definedas \left\{\boldl \in \Z_+^n \text{ such that }
\sum_{i=1}^n 2^{-l_{i}} \leq 1\right\}.$$ 

Because Huffman's algorithm\cite{Huff} finds codes minimizing average redundancy (\ref{minavg}), the \textit{minimum-average redundancy problem} itself is often referred to as the ``\textit{Huffman problem},'' even though the problem did not originate with Huffman himself.  Huffman's algorithm is a greedy algorithm built on the observation that the two least likely items will have the same length and can thus be considered siblings in the coding tree.  A reduction is thus made in which the two items of weights $w(i)$ and $w(j)$ are considered as one with combined weight $w(i)+w(j)$.  The codeword of the combined item determines all but the last bit of each of the items combined, which are differentiated by this last bit.  This reduction continues until there is one item left, and, assigning this item the null string, a code is defined for all input items.  In the corresponding optimal code tree, the $i$\textsuperscript{th} leaf corresponds to the codeword of the $i$\textsuperscript{th} input item, and thus has weight $w(i)$, whereas the weight of parent nodes are determined by the combined weight of the corresponding merged item.  

We began by stating that an optimal $\boldl^\opt$ must satisfy
$$H(\p) \leq \sum_{i \in \X} p_il_i^\opt < H(\p) + 1$$
or, equivalently,
$$0 \leq \bar{R}(\boldl^\opt,\p) < 1.$$
Less well known is that simple changes to the Huffman algorithm solve
several related coding problems which optimize for different
objectives.  We discuss three such problems, all three of
which have been previously shown to satisfy redundancy bounds for optimal
$\tilde{\boldl}$ of the form
$$\tilde{H}(\p) \leq \tilde{L}(\p,\tilde{\boldl}) < \tilde{H}(\p)+1$$
or 
$$0 \leq \tilde{R}(\tilde{\boldl},\p) < 1$$ for some entropy measure
$\tilde{H}$ and cost measure $\tilde{L}$, or for some redundancy measure
$\tilde{R}$.

In this paper, we improve these bounds in a similar manner to
improvements made to the Huffman problem: Given $p_1$, the probability
of the most likely item, the Huffman problem improvements find
functions $\bar{o}(p_1)$ and/or $\bar{\omega}(p_1)$ such that
$$0 \leq \bar{o}(p_1) \leq \bar{R}(\boldl^\opt,\p) \leq
\bar{\omega}(p_1) < 1.$$ The smallest $\bar{\omega}$, tight over most
$p_1$, is given in \cite{Mans}, while a tight $\bar{o}$ is given in
\cite{MoAb}.  Tight bounds given any value $p_j$ in $\p$, would yield
alternative functions $\bar{o}'(p_j)$ and $\bar{\omega}'(p_j)$ such that
$$0 \leq \bar{o}'(p_j) \leq \bar{R}(\boldl^\opt,\p) \leq
\bar{\omega}'(p_j) < 1.$$  In this case, tight bounds are given by
\cite{MPK}, which also addresses lower bounds given the
\textit{least} probable symbol, which we do not consider here.

In the following, we wish to find functions $\tilde{o}$,
$\tilde{\omega}$, $\tilde{o}'$, and/or $\tilde{\omega}'$
such that
$$0 \leq \tilde{o}(p_1) \leq \tilde{R}(\tilde{\boldl},\p) \leq \tilde{\omega}(p_1) \leq 1$$ and/or
$$0 \leq \tilde{o}'(p_j) \leq \tilde{R}(\tilde{\boldl},\p) \leq \tilde{\omega}'(p_j) \leq 1$$
in the case of redundancy objectives, 
and to find 
$\Tilde{\Tilde{o}}$, $\Tilde{\Tilde{\omega}}$,
$\Tilde{\Tilde{o}}'$, and/or $\Tilde{\Tilde{\omega}}'$ such that
$$0 \leq \Tilde{\Tilde{o}}(\tilde{H}(\p), p_1) \leq \tilde{L}(\p,\tilde{\boldl})\leq \Tilde{\Tilde{\omega}}(\tilde{H}(\p), p_1) \leq 1$$ and/or
$$0 \leq \Tilde{\Tilde{o}}'(\tilde{H}(\p), p_j) \leq \tilde{L}(\p,\tilde{\boldl})\leq \Tilde{\Tilde{\omega}}'(\tilde{H}(\p), p_j) \leq 1$$
in the case of other length objectives.

All of the nonlinear objectives we consider have been shown to be
solved by generalized versions
of the Huffman algorithm\cite{HKT,Park,Knu1,ChTh,Baer05}.  These
generalizations change the combining rule; instead of replacing items
$i$ and $j$ with an item of weight $w(i)+w(j)$, the generalized
algorithm replaces them with an item of weight $f(w(i),w(j))$ for some
function~$f$.  The weight of a combined item (a node) therefore need
not be equal to the sum of the probabilities of the items merged to
create it (the sum of the leaves of the corresponding subtree).  Thus
the sum of weights in a reduced problem need not be~$1$, unlike in the
original Huffman algorithm.  In particular, the weight of the root,
$w_\root$, need not be~$1$.  However, we continue to assume that the
sum of \textit{inputs} to the coding problems will be~$1$ (with the exception
of reductions among problems).

The next section introduces the objectives of interest, along with their motivations and our main contributions.  These contributions, indicated by (\lefthand), are bounds on performance of optimal codes according to their optimizing objectives, as well as related properties.  We defer the formal presentation of these contributions, along with proofs, until later sections, where they are presented as theorems and corollaries, along with the remarks immediately following them and associated figures.  These begin in Section~\ref{bred}, where we find tight exhaustive bounds for the values of minimized maximum pointwise redundancy (\ref{mmprDef}) and corresponding $l_j$ in terms of~$p_j$.  Pointwise redundancy for a symbol $i$ is $l_i + \lg p_i$.  In Section~\ref{dthbounds}, we then extend these to exhaustive --- but not tight --- bounds for minimized $d$\textsuperscript{th} exponential redundancy (\ref{dth}), a measure which takes a $\beta$-exponential average\cite{AcDa} of pointwise redundancy (where, in this case, parameter $\beta$ is~$d$).  In Section~\ref{bexp}, we investigate the behavior of codes with minimized exponential average (\ref{one}), including bounds and optimizing $l_1$ in terms of~$p_1$.

\section{Objectives, Motivations, and Main Results}

\subsection{Maximum pointwise redundancy ($R^*$)}

The most recently proposed problem objective we consider is that
formulated by Drmota and Szpankowski\cite{DrSz}.  Instead of
minimizing average redundancy $\bar{R}(\boldl,\p) \definedas \sum_{i
  \in \X} p_i (l_{i}+\lg p_i)$, here we minimize maximum pointwise
redundancy
\begin{equation}
R^*(\boldl,\p) \definedas \max_{i \in \X} (l_{i}+\lg p_i).
\label{mmprDef}
\end{equation}
An extension of Shannon coding introduced by Blumer and
McEliece\cite[p.~1244]{BlMc} to upper-bound the problem considered in
Section~\ref{expavg} was later rediscovered and efficiently implemented
by Drmota and Szpankowski as a solution to this maximum pointwise
redundancy problem.  A subsequent solution to the problem is a
variation of Huffman coding\cite{Baer05} derived from that in
\cite{Golu}, one using combining rule
\begin{equation}
f^*(w(i),w(j)) \definedas 2\max(w(i),w(j)).
\label{mmprcomb}
\end{equation}

\subsubsection*{\textbf{Applications in prior literature}}
The solution of this worst-case pointwise redundancy problem is relevant to optimizing maximal (worst-case) minimax redundancy, a universal modeling problem (as in \cite[p.~176]{Shta}) for which the set $\setP$ of possible probability distributions results in a normalized ``maximum likelihood distribution.''\cite{DrSz}  More recently Gawrychowski and Gagie proposed a second worst-case redundancy problem which also finds its solution in minimizing maximum pointwise redundancy\cite{GaGa}.  For this problem, normalization is not relevant and one allows any probability distribution that is consistent with an empirical distribution based on sampling.

\subsubsection*{\textbf{Prior and current results}}

The first proposed algorithm for the maximum pointwise redundancy problem is a codeword-wise improvement on the Shannon code in the sense that each codeword is the same length as or one bit shorter than the corresponding codeword in the Shannon code.  This method is called ``generalized Shannon coding.''  (With proper tie-breaking techniques, the Huffman-like solution guarantees that each codeword, in turn, is no longer than the generalized Shannon codeword.  As both methods guarantee optimality, this makes no difference in the maximum pointwise redundancy.)  Notice a property true of Shannon codes --- generalized or not --- but not minimum average redundancy (Huffman) codes: Because any given codeword has a length $l_i$ not exceeding $\lceil -\lg p_i \rceil$, this length is within one bit of the associated input symbol's self-information, $-\lg p_i$.  This results in bounds of $R_\opt^*(\p) \in [0,1)$, which are improved upon in Section~\ref{bred}.  The bound can also be considered a degenerative case from a result of Shtarkov \cite[p.~176]{Shta}, that for which the probabilities are fully known.

\begin{figure}[t]
     \psfrag{R}{\Large $R_\opt^*(\p)$}
     \psfrag{p(1)}{\Large $p_1$ or $p_j$}
     \centering
          \resizebox{8.75cm}{!}{\includegraphics{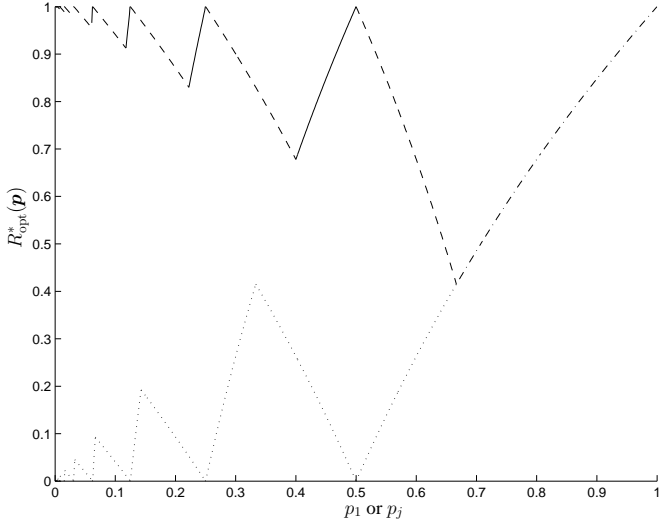}}
     \caption{Tight bounds on minimum maximum pointwise redundancy,
     including achievable upper bounds (solid), approachable upper
     bounds (dashed), achievable lower bounds (dotted), and fully
     determined values for $p_1 = p_j \geq 2/3$ (dot-dashed).}
     \label{mmprComplete}
\end{figure}

The aforementioned papers contain further analysis of this problem,
but no improved closed-form bounds of the type introduced here.  Here
results are given as strict upper and lower bounds in
Theorem~\ref{mmprbetter} in Section~\ref{bred1}.  Specifically, whether 
considering known $p_j$ in general or known $p_1$ in particular, this
problem has upper bound
\begin{equation}
\omega'^*(p_j) = \max\left(1+\lg \frac{1-p_j}{1-2^{-\lu_j}}, \lu_j+\lg p_j\right)
\tag{\lefthand}
\end{equation}
where this and $\omega^*(p_1) = \omega'^*(p_1)$ are tight bounds.  Also, it has
lower bound
\begin{equation}
o'^*(p_j) = \min\left(\lu_j+\lg p_j, \lg \frac{1-p_j}{1-2^{-\lu_j+1}}\right)
\tag{\lefthand}
\end{equation}
for $p_j < 0.5$, and $1+\lg p_j$ otherwise, and, again, this and
$o'^*(p_j) = o^*(p_j)$ are tight.  Here $\lu_j \definedas \lceil - \lg
p_j \rceil$, the results are illustrated in Fig.~\ref{mmprComplete},
and the values for which the maximum and minimum apply are given in
the theorem (i.e., the bounds are tight).

Further results here include those regarding codeword length;
Theorem~\ref{mmprlen} states that any optimal code will have $l_j
\leq \nu$ if $p_j \geq 2^{-\nu}$ and that any probability distribution
with $p_j \leq 1/(2^{\nu}-1)$ will be optimized by at least one code
with $l_j \geq \nu$.

\subsection{$d$\textsuperscript{th} exponential redundancy ($R^d$)}

A spectrum of problems bridges the objective of Huffman coding with the objective optimized by generalized Shannon coding using an objective proposed in \cite{Nath} and solved for in \cite{Park}.  In this particular context, the range of problems, parameterized by a variable $d$, can be called $d$\textsuperscript{th} exponential redundancy\cite{Baer05}.  Such problems involve the minimization of the following:
\begin{equation} 
R^d(\p,\boldl) \definedas \frac{1}{d} \lg \sum_{i \in \X} p_i^{1+d} 2^{dl_{i}} =
\frac{1}{d} \lg \sum_{i \in \X} p_i 2^{d(l_{i}+\lg p_i)}.
\label{dth} 
\end{equation}
Although positive parameter $d$ is the case we consider most often
here, $d \in (-1,0)$ is also a valid minimization problem.  If we let $d
\rightarrow 0$, we approach the average redundancy (Huffman's
objective), while $d \rightarrow \infty$ is maximum pointwise
redundancy\cite{Baer05}.  The combining rule, introduced in 
\cite[p.~486]{Park}, is
\begin{equation}
f^d(w(i),w(j)) \definedas \left(2^d w(i)^{1+d}+2^d w(j)^{1+d}\right)^{\frac{1}{1+d}} .
\label{dhuff}
\end{equation}

\subsubsection*{\textbf{Prior and current results and applications}}

This redundancy objective is less analyzed than the others mentioned
here, likely because there are no direct applications in the published
literature.  However, it is closely related not only to average
redundancy and to maximum pointwise redundancy, but also to the more
applicable objective considered in Section~\ref{expavg}.  Solution
properties of these objectives --- including redundancy bounds --- can
therefore be related via $d$\textsuperscript{th} exponential
redundancy.  In particular, as we show in Section~\ref{dthbounds}, the
upper bound for maximum pointwise redundancy also improves upon the
already-known bound for $d$\textsuperscript{th} exponential
redundancy,
$$R_\opt^d(\p) \definedas \min_{\boldsl \in \L} R_\opt^d(\boldl,\p) \in [0,1).$$
Given $d > 0$, we show in
Corollary~\ref{dcorollary} that any upper bound on minimax pointwise
redundancy and any lower bound on minimum average redundancy serve to
bound $d$\textsuperscript{th} exponential redundancy.

Specifically, consider $\omega^*$ given above (identical to
$\omega'^*$) and lower bound on minimum average redundancy $\bar{o}$
given in the literature\cite{MoAb} (identical to
$\bar{o}'$\cite{MPK}).  For any probability $p_j$ in input
distribution $\p$ and any $d > 0$,
\begin{equation}
\bar{o}(p_j) \leq R_\opt^d(\p) \leq \omega^*(p_j)
\tag{\lefthand}
\end{equation}
as illustrated in Fig.~\ref{dar} and detailed in the corollary.
Furthermore, any upper bound on minimum
average redundancy --- e.g., $\bar{\omega}(p_1)$\cite{Mans} or 
$\bar{\omega}'(p_j)$\cite{YeYe2,MPK} --- similarly bounds
$d$\textsuperscript{th} exponential redundancy with $d \in (-1,0)$.

\subsection{Exponential average ($\Cost_\q$)}
\label{expavg}

A related problem is one proposed by Campbell\cite{Camp0, Camp}.  This exponential problem, given probability mass function $\p$ and $\q \in (0,\infty) \backslash 1$, is to find a code minimizing
\begin{equation} 
\Cost_\q(\p,\boldl) \definedas \log_\q \sum_{i \in \X} p_i \q^{l_{i}} .
\label{one} 
\end{equation}
In this case our parameter $\q$ is the base, rather than the exponential scaling factor, although much prior work does express this problem in the equivalent alternative form,
$$
\Cost_\q(\p,\boldl) = \frac{1}{(\lg \q)} \lg \sum_{i \in \X} p_i 2^{(\lg \q)l_i} .
$$
The solution to this \cite{HKT, Park, Humb2} uses combining rule 
\begin{equation}
f_\q(w(i),w(j)) \definedas \q w(i)+\q w(j).
\label{expcomb} 
\end{equation}
A change of variables transforms the $d$\textsuperscript{th}
exponential redundancy problem into (\ref{one}) by assigning $\q = \lg
d$ and using input weights $w(i)$ proportional to $p_i^{1+d}$, which
yields~(\ref{dhuff}).  We illustrate this precisely in
Section~\ref{dthbounds} in (\ref{trans}), which we use in
Section~\ref{bexp} to find initial improved entropy bounds.  These are
supplemented by additional bounds for problems with $\q \in (0.5,1)$
and $p_1 \geq 2\q/(2\q+3)$ (as illustrated in Fig.~\ref{p1}
at the end of Section~\ref{bexp}).

\subsubsection*{\textbf{Applications} ($\q < 1$)}

It is important to note here that $\q > 1$ is an average of growing
exponentials, while $\q < 1$ is an average of decaying exponentials.
These two subproblems have different properties and
have often been considered separately in the literature.  An
application for the decaying exponential variant involving single-shot
communications has a communication channel with a window of
opportunity of a total duration (in bits) distributed geometrically
with parameter~$\q$ \cite{Baer07}.  The probability of successful
transmission is
\begin{equation}
\P[\mbox{success}] = \q^{\Cost_\q(\sp,\boldsl)} = \sum_{i=1}^n p_i \q^{l_i}.
\label{success}
\end{equation}
For $\q > 0.5$, the unit-sized bound we improve upon is in terms of
R\'{e}nyi entropy, as in (\ref{LH01}); the solution is trivial for
$\q \leq 0.5$, as we note at the start of Section~\ref{bexp}.

\subsubsection*{\textbf{Applications} ($\q > 1$)}

We add an additional observation on a modified version of this
problem: Suppose there are a sequence of windows of opportunities
rather than only one.  The probability that a window stays open long
enough to send a message of length $l_i$ is $\q^{l_i}$, since each
additional bit has independent probability $\q$ of getting through.
Thus, given $l_i$, the expected number of windows needed to send a
message --- assuming it is necessary to resend communication for each
window --- is the multiplicative inverse of this.  Overall expectation
is therefore
$$\E[N] = \sum_{i=1}^n p_i \q^{-l_i} = \q^{-\Cost_{\q^{-1}}(\sp,\boldsl)} .$$
Although such a resending of communications is usually not needed for
a constant message, this problem is a notable dual to the first
problem.  In this dual problem, we seek to minimize the expectation of
a growing exponential of lengths rather than maximize the expectation
of a decaying exponential.

Originally, the $\q>1$ variation of (\ref{one}) was used in Humblet's
dissertation\cite{Humb0} for a queueing application originally
proposed by Jelinek\cite{Jeli} and expounded upon in \cite{BlMc}.
This problem is one in which overflow probability should be minimized,
where the source produces symbols randomly and the codewords are
temporarily stored in a finite buffer.  In this problem, there exists an
$\q > 1$ such that optimizing $\Cost_\q(\p,\boldl)$ optimizes this problem;
the correct $\q$ is found through iteration.  The Huffman-like coding method
was simultaneously published in \cite{HKT, Park, Humb2}; in the last
of these, Humblet noted that the Huffman combining method
(\ref{expcomb}) finds the optimal code with $\q \in (0,1)$ as well.

More recently, the $\q > 1$ variation was shown to have a third
application\cite{ReCh}.  In this problem, the true probability of the
source is not known; it is only known that the relative entropy
between the true probability and $\p$ is within some known bound.  As
in Humblet's queueing problem, there is an $\q > 1$, found via
iteration, such that optimizing $\Cost_\q(\p,\boldl)$ solves the
problem.

\subsubsection*{\textbf{Prior and current results}}

Note that $\q \rightarrow 1$ corresponds to the usual linear expectation objective.  Problems for $\q$ near $1$ are of special interest, since $\q \downarrow 1$ corresponds to the minimum-variance solution if the problem has multiple solutions --- as noted in \cite{Park}, among others --- while $\q \uparrow 1$ corresponds to the maximum-variance solution.

Most of the aforementioned improved bounds are based on a given
highest symbol probability,~$p_1$.  We thus give this case special
attention and also discuss the related property of the length of the
most likely codeword in these coding problems.  The bounds in this
paper are the first of their kind for nontraditional Huffman codes,
bounds which are, for $\Cost_\q$, functions of both entropy and $p_1$,
as in the traditional case.  However, they are not the first improved
bounds for such codes.  More sophisticated bounds on the optimal
solution for the exponential-average objective are given in
\cite{BlMc} for $\q>1$; these appear as solutions to related problems
rather than in closed form, however, and these problems require no
less time or space to solve than the original problem.  They are
mainly useful for analysis.  Bounds given elsewhere for a closely
related objective having a one-to-one correspondence with (\ref{one})
are demonstrated under the assumption that $p_1 \geq 0.4$ always
implies $l_{1}$ can be $1$ for the optimal code\cite{Tane}.  We show
that this is not necessarily the case due to the difference between
the exponential-average objective and the usual objective of an
arithmetic average.

Specifically, Theorem~\ref{l1} states that, for $\q \in (0.5,1]$, a
code with shortest codeword of length $1$ is optimal if $p_1 \geq
2\q/(2\q+3)$.  Furthermore, for $\q > 1$, no value of $p_1 \in (0,1)$
guarantees $l_1 = 1$, and, for $\q \leq 0.5$, there is always an
optimal code with $l_1 = 1$, regardless of the input distribution.
This results in the improved bounds of Corollary~\ref{coro}; when
$\q \in (0.5,1)$ and $p_1 \geq 2\q/(2\q+3)$, optimal $\boldl$
satisfies
\begin{equation}
\begin{array}{l}
\displaystyle \q^2 \left[ \q^{{\alpha} H_{\alpha}(\sp)} - p_1^{\alpha} \right]^{\frac{1}{\alpha}} + \q p_1 < \left(\sum_{i=1}^n p_i \q^{l_{i}}\right) \\
\displaystyle \qquad \leq \q \left[\q^{{\alpha} H_{\alpha}(\sp)} - p_1^{\alpha} \right]^{\frac{1}{\alpha}} + \q p_1
\end{array}
\tag{\lefthand}
\end{equation}
where $\alpha = 1/(1+\lg \q)$, and R\'{e}nyi entropy 
$$H_\alpha(\p) \definedas \frac{1}{1-\alpha}\lg \sum_{i=1}^n p_i^\alpha.$$
This is an improvement on the unit-sized bounds, $$H_{\alpha}(\p) \leq \log_\q
\sum_{i \in \X} p_i \q^{l_{i}} < H_{\alpha}(\p) + 1.$$ In addition, we
show in Corollary~\ref{ecorollary} that a reduction from this problem
to $d$\textsuperscript{th} exponential redundancy extends the
nontrivial bounds for the redundancy utility to nontrivial bounds for
any $\q > 0.5$, resulting in
\begin{equation}
0 \leq \Cost_\q^\opt(\p) - H_{\alpha}(\p) \leq \bar{\omega}\left(p_1^{\alpha} 2^{(\alpha-1)H_{\alpha}(\sp)}\right)
\tag{\lefthand}
\end{equation}
and
\begin{equation}
0 \leq \Cost_\q^\opt(\p) - H_{\alpha}(\p) \leq \bar{\omega}'\left(p_j^{\alpha} 2^{(\alpha-1)H_{\alpha}(\sp)}\right)
\tag{\lefthand}
\end{equation}
for $\q \in (0.5,1)$ and
\begin{equation}
\begin{array}{rcl}
\displaystyle \bar{o}\left(p_j^{\alpha} 2^{(\alpha-1)H_{\alpha}(\sp)}\right) 
&\leq& \displaystyle \Cost_\q^\opt(\p) - H_{\alpha}(\p) \\
&\leq& \displaystyle \omega^*\left(p_j^{\alpha} 2^{(\alpha-1)H_{\alpha}(\sp)}\right)
\end{array}
\tag{\lefthand}
\end{equation}
for $\q > 1$.
 
\section{Maximum Pointwise Redundancy}
\label{bred}

\subsection{Maximum pointwise redundancy bounds}
\label{bred1}

Shannon found redundancy bounds for $\bar{R}_\opt(\p)$, the average redundancy $\bar{R}(\boldl,\p) = \sum_{i \in \X} p_i l_{i} - H(\p)$ of the average redundancy-optimal $\boldl$.  The simplest bounds for minimized maximum pointwise redundancy
$$R_\opt^*(\p) \definedas \min_{\boldsl \in \L} \max_{i \in \X}
\left(l_{i}+\lg p_i\right)$$ are quite similar to and can be combined with 
Shannon's bounds as follows:
\begin{equation}
0 \leq \bar{R}_\opt(\p) \leq R_\opt^*(\p) < 1
\label{mmprbounds}
\end{equation}
The average redundancy case is a lower bound  because the maximum ($R^*(\boldl,\p)$) of the values ($l_{i}+\lg p_i$) that average to a quantity ($\bar{R}(\boldl,\p)$) can be no less than the average (a fact that holds for all $\boldl$ and~$\p$).  The upper bound is due to Shannon code $l_i^0(\p) \definedas \lceil -\lg p_i \rceil$ resulting in $$R_\opt^*(\p) \leq R^*(\boldl^0(\p),\p) = \max_{i \in \X}{(\lceil -\lg p_i \rceil + \lg p_i)} < 1.$$

A few observations can be used to find a series of improved lower and
upper bounds on optimum maximum pointwise redundancy based on
(\ref{mmprbounds}):

\subsubsection*{\textbf{Properties, Maximum Pointwise Redundancy}}
\begin{lemma}
Suppose we apply (\ref{mmprcomb}) to find a Huffman-like code tree in
order to minimize maximum pointwise redundancy ($\bar{R}(\boldl,\p)$ 
given $\p$).  Then the following holds:
\begin{enumerate}
\item Items are always merged by nondecreasing weight.
\item The weight of the root $w_\root$ of the coding tree
determines the maximum pointwise redundancy, $R^*(\boldl,\p) = \lg
w_\root$.
\item The total probability of any subtree is no greater than the
total weight of the subtree.  
\item If $p_1 \leq 2p_{n-1}$, then a minimum maximum pointwise
redundancy code can be represented by a \textit{complete tree}, that is, a tree
with leaves at depth $\lfloor \lg n \rfloor$ and $\lceil \lg n \rceil$
only (with $\sum_{i \in \X} 2^{-l_{i}} = 1$).  (This property is similar to the property noted in \cite{GaVV} for optimal-expected-length codes of sources termed \textit{quasi-uniform} in \cite{MSW}.)
\end{enumerate}
\label{complete}
\end{lemma}

\begin{proof}
We use an inductive proof in which base cases of sizes $1$ and $2$ are
trivial, and we use weight function $w$ instead of probability mass
function~$\p$ to emphasize that the sums of weights need not
necessarily add up to~$1$.  Assume first that all properties here are
true for trees of size $n-1$ and smaller.  We wish to show that they
are true for trees of size~$n$.

The first property is true because
$f^*(w(i),w(j))=2\max(w(i),w(j))>w(i)$ for any $i$ and~$j$; that is, a
compound item always has greater weight than either of the items
combined to form it.  Thus, after the first two weights are combined,
all remaining weights, including the compound weight, are no less than
either of the two original weights.

Consider the second property.  After merging the two least weighted of
$n$ (possibly merged) items, the property holds for the resulting
$n-1$ items.  For the $n-2$ untouched items, $l_{i}+\lg w(i)$ remains	
the same.  For the two merged items, let $l_{n-1}$ and $w(n-1)$ denote
the maximum depth/weight pair for item $n-1$ and $l_{n}$ and $w(n)$ the
pair for $n$.  If $l'$ and $w'$ denote the depth/weight pair of the
combined item, then 
\begin{eqnarray*}
l'+\lg w' &=& l_{n} - 1 + \lg (2 \max(w(n-1),w(n))) \\
&=& \max(l_{n-1} + \lg w(n-1),l_{n} + \lg w(n)).
\end{eqnarray*}
Thus the two trees have
identical maximum redundancy, which is equal to $\lg w_\root$ since
the root node is of depth~$0$.  Consider, for example, $p =
(0.5,0.3,0.2)$, which has optimal codewords with lengths $\boldl =
(1,2,2)$.  The first combined pair has 
\begin{eqnarray*}
l'+\lg w' = 1+\lg 0.6 &=& \max(2+\lg 0.3,2+\lg 0.2)\\
&=& \max(l_{2}+\lg p_2,l_{3}+\lg p_3).
\end{eqnarray*}
This value is identical to that of the maximum redundancy, $\lg 1.2 =
\lg w_\root$.

For the third property, the first combined pair yields a weight that
is no less than the combined probabilities.  Thus, via induction, the
total probability of any (sub)tree is no greater than the weight of
the (sub)tree.

In order to show the final property, first note that $\sum_{i \in \X}
2^{-l_{i}} = 1$ for any tree created using the Huffman-like procedure,
since all internal nodes have two children.  Now think of the
procedure as starting with a (priority) queue of input items, ordered by
nondecreasing weight from head to tail.  After merging two items,
obtained from the head of the queue, into one compound item, that item
is placed back into the queue as one item, but not necessarily at the
tail; an item is placed such that its weight is no smaller than any
item ahead of it and is smaller than any item behind it.  In keeping
items ordered, this results in an optimal coding tree.  A variant of
this method can be used for linear-time coding\cite{Baer05}.

In this case, we show not only that an optimal complete tree exists,
but that, given an $n$-item tree, all items that finish at level
$\lceil \lg n \rceil$ appear closer to the head of the queue than any
item at level $\lceil \lg n \rceil - 1$ (if any), using a similar
approach to the proof of Lemma~2 in \cite{Baer07}.  Suppose this is
true for every case with $n-1$ items for $n>2$, that is, that all
nodes are at levels $\lfloor \lg (n-1) \rfloor$ or $\lceil \lg (n-1)
\rceil$, with the latter items closer to the head of the queue than
the former.  Consider now a case with $n$ nodes.  The first step of
coding is to merge two nodes, resulting in a combined item that is
placed at the end of the combined-item queue, as we have asserted that
$p_1 \leq 2p_{n-1} = 2\max(p_{n-1},p_n)$.  Because it is at the end
of the queue in the $n-1$ case, this combined node is at level
$\lfloor \lg (n-1) \rfloor$ in the final tree, and its children are at
level $1+\lfloor \lg (n-1) \rfloor = \lceil \lg n \rceil$.  If $n$ is
a power of two, the remaining items end up on level $\lg n = \lceil
\lg (n-1) \rceil$, satisfying this lemma.  If $n-1$ is a power of two,
they end up on level $\lg (n-1) = \lfloor \lg n \rfloor$, also
satisfying the lemma.  Otherwise, there is at least one item ending up
at level $\lceil \lg n \rceil = \lceil \lg (n-1) \rceil$ near the head
of the queue, followed by the remaining items, which end up at level
$\lfloor \lg n \rfloor = \lfloor \lg (n-1) \rfloor$.  In any case, all
properties of the lemma are satisfied for $n$ items, and thus for any
number of items.
\end{proof}

We can now present the improved redundancy bounds.

\subsubsection*{\textbf{Bounds, Maximum Pointwise Redundancy}}
\begin{theorem}
For any distribution in which there exists a $p_j \geq 2/3$,
$R_\opt^*(\p) = 1+\lg p_j$.  If $p_j \in [0.5,2/3)$, then $R_\opt^*(\p) \in [1+\lg
p_j,2+\lg (1-p_j))$ and these bounds are tight not only for general $p_j$, but
for~$p_1$, in the sense that we can find probability mass functions with
the given $p_1 = p_j$ achieving the lower bound and approaching the upper
bound.  Define $\lu_j \definedas \lceil - \lg p_j \rceil$.  Thus $\lu_j$
satisfies $p_j \in [2^{-\lu_j}, 2^{-\lu_j+1})$, and $\lu_j > 1$ for
$p_j \in (0,0.5)$; in this range, the following bounds for $R_\opt^*(\p)$ 
are tight for general $p_j$ and $p_1$ in particular:
$$
\begin{array}{ll}
\quad p_j & \quad R_\opt^*(\p) \\
\hline  
& \\[-5pt]
\left[\frac{1}{2^{\lu_j}},\frac{1}{2^{\lu_j}-1}\right) &
\left[\lu_j+\lg p_j,1+\lg \frac{1-p_j}{1-2^{-\lu_j}}\right) \\[6pt]
\left[\frac{1}{2^{\lu_j}-1},\frac{2}{2^{\lu_j}+1}\right) &
\left[\lg \frac{1-p_j}{1-2^{-\lu_j+1}},1+\lg \frac{1-p_j}{1-2^{-\lu_j}}\right) \\[6pt]
\left[\frac{2}{2^{\lu_j}+1},\frac{1}{2^{\lu_j-1}}\right) &
\left[\lg \frac{1-p_j}{1-2^{-\lu_j+1}},\lu_j+\lg p_j \right]\\
~&~\\
\end{array}
$$
\label{mmprbetter}
\end{theorem}

\begin{proof}
The key here is generalizing the unit-sized bounds of (\ref{mmprbounds}).

\subsubsection{Upper bound}
Before we prove the upper bound, note that, once proven, the tightness
of the upper bound in $[0.5,1)$ is shown via
$$\p = \left(p_j,1-p_j-\epsilon,\epsilon\right)$$ for which the bound
is achieved in $[2/3,1)$ for any $\epsilon \in (0,(1-p_j)/2]$ and
approached in $[0.5,2/3)$ as $\epsilon \downarrow 0$. 

Let us define what we call a \textit{$j$-Shannon code}:
$$
l_i^j(\p) = \left\{
\begin{array}{ll}
\lu_j \definedas \left\lceil -\lg p_j \right\rceil,& i = j \\ 
\left\lceil -\lg
\left(p_i\left(\frac{1-2^{-\lu_j}}{1-p_j}\right)\right) \right\rceil,& i \neq j
\end{array}
\right.
$$ This code was previously presented in the context of finding
\textit{average} redundancy bounds given any probability \cite{YeYe2}.
Here it improves upon the original Shannon code $\boldl^0(\p)$ by
making the length $l_j^j$ of the $j$\textsuperscript{th} known
codeword $\lambda_j$, and taking this length into account when
designing the rest of the code.  The code satisfies the Kraft
inequality, and thus, as a valid code, its redundancy is an upper
bound on the redundancy of an optimal code.  Note that
\begin{eqnarray}
\lefteqn{\max_{i \neq j} (l_i^j(\p) + \lg p_i)} \nonumber \\
&=& \max_{i \neq j} \left(
\left\lceil \lg \frac{1-p_j}{p_i(1-2^{-\lu_j})} \right\rceil
+ \lg p_i\right) \\
&<& 1+\lg \frac{1-p_j}{1-2^{-\lu_j}}. \nonumber
\label{firstorder}
\end{eqnarray}
There are two cases:

\paragraph{$p_j \in [2/(2^{\lu_j}+1),1/2^{\lu_j-1})$}
In this case, the maximum pointwise redundancy of the item $j$ in code
$l^j(\p)$ is no less than $1+\lg ((1-p_j)/(1-2^{-\lu_j}))$.  Thus, due to
(\ref{firstorder}),
$$R_\opt^*(\p) \leq R^*(\boldl^j(\p),\p) = \lu_j+\lg p_j.$$  If $\lu_j > 1$ 
and $p_j \in [2/(2^{\lu_j}+1),1/2^{\lu_j-1})$, consider $j = 1$ and probability 
mass function
$$\p = \left(p_1,\underbrace{\frac{1-p_1-\epsilon}{2^{\lu_1}-2},
\ldots,\frac{1-p_1-\epsilon}{2^{\lu_1}-2}}_{2^{\lu_1}-2}, \epsilon\right)$$
where $\epsilon \in (0, 1-p_12^{\lu_1-1})$.  Because $p_1 \geq 2/(2^{{\lu_1}}+1)$, 
$$1-p_12^{{\lu_1}-1} \leq (1-p_1-\epsilon)/(2^{\lu_1}-2)$$ 
and $p_{n-1} \geq p_n$.  Similarly,
$p_1 < 1/2^{{\lu_1}-1}$ assures that $p_1 \geq p_2$, so the
probability mass function is monotonic.  Since $2p_{n-1} > p_1$, 
by Lemma~\ref{complete}, an optimal code for this probability mass
function is $l_i = \lu_1$ for all~$i$, achieving $R^*(\boldl,\p) = \lu_1
+ \lg p_1$.   Since $j=1$ has the maximum pointwise redundancy,
this upper bound is tight whether considering $p_1$ or general~$p_j$.

\paragraph{$p_j \in [1/2^{\lu_j}, 2/(2^{\lu_j}+1))$}
In this case, (\ref{firstorder}) immediately results in $$R_\opt^*(\p)
\leq R^*(\boldl^j(\p),\p) < 1+\lg ((1-p_j)/(1-2^{-\lu_j})).$$
Again considering $j = 1$, the probability mass function
$$\p = \left(p_1,\underbrace{\frac{1-p_1-\epsilon}{2^{\lu_1}-1},
  \ldots,\frac{1-p_1-\epsilon}{2^{\lu_1}-1}}_{2^{\lu_1}-1}, \epsilon\right)$$
illustrates the tightness of this bound for $\epsilon \downarrow 0$.
For sufficiently small~$\epsilon$, this probability mass function is
monotonic and $p_1 < 2p_{n-1}$.  Lemma~\ref{complete} then indicates
that an optimal code has $l_{i} = \lu_1$ for $i \in \{1, 2, \ldots,
n-2\}$ and $l_{n-1}=l_{n}=\lu_1+1$.  Thus the bound is approached with
item $n-1$ having the maximum pointwise redundancy.

\subsubsection{Lower bound}

Here we first address the lower bound given~$p_1$. 
Consider all optimal codes with $l_{1} = \mu$ for
some fixed $\mu \in \{1, 2, \ldots\}$.  If $p_1 \geq 2^{-\mu}$,
$R^*(\boldl,\p) \geq l_{1} + \lg p_1 = \mu + \lg p_1$.  If $p_1 <
2^{-\mu}$, consider the weights at level $\mu$ (i.e., $\mu$ edges
below the root).  One of these weights is $p_1$, while the rest are
known to sum to a number no less than $1-p_1$.  Thus at least one
weight must be at least $(1-p_1)/(2^{\mu}-1)$ and $R^*(\boldl,\p) \geq
\mu + \lg ((1-p_1)/(2^{\mu}-1))$.  Thus,
$$R_\opt^*(\p) \geq \mu + \lg \max\left(p_1,\frac{1-p_1}{2^{\mu}-1}\right)$$
for $l_1 = \mu$, and since this can be any positive integer,
$$R_\opt^*(\p) \geq \min_{\mu \in \{1,2,3,\ldots\}} \left(\mu + \lg
\max\left(p_1, \frac{1-p_1}{2^{\mu}-1}\right)\right)$$ which is equivalent
to the bounds provided.

For arbitrary $p_j$, the approach is similar, but a modification is
required to the above when $p_j < 2^{-\mu}$; we are no longer
guaranteed to have $2^\mu$ nodes on level $\mu$ (where $\mu = l_j$).
Instead, consider the set of leaves $\setA$ above level $\mu$ and the
set of nodes $\setN$ on level~$\mu$.  Let $\setA'$ be a set of nodes
(not actually in the optimal tree) such that, for each leaf $i$ in
$\setA$, there are $2^{\mu-l_i}$ nodes in $\setA'$, each one having
weight $p_i2^{l_i-\mu}$.  Thus the combined probability of $\setA'$
remains the same and the combined weight of $\setA'$ and $\setN$ is no
less than $1$.  The cardinality of the combined sets --- which can be
considered as the level of an extended tree --- is $2^\mu$, for the
same reason that this is the number of nodes on the level for the case
of known~$p_1$.  Thus the maximum weight of the $2^\mu-1$ items in
$\setA' \cup \setN \backslash \{j\}$ is at least its average, which
is, in turn, at least $(1-p_j)/(2^{\mu}-1)$.  If that item is in
$\setN$, it follows, as above, that $R^*(\boldl,\p) \geq \mu + \lg
((1-p_j)/(2^{\mu}-1))$.  If it is not, then it --- along with
$2^{\mu-l_i}$ items of the same weight --- corresponds to an item $i$
with $p_i \geq (1-p_j)/(2^{l_i}-2^{l_i-\mu})$.  In this case, too,
\begin{eqnarray*}
R^*(\boldl,\p) &\geq& 
l_i + \lg ((1-p_j)/(2^{l_i}-2^{l_i-\mu})\\
&=& \mu + \lg ((1-p_j)/(2^{\mu}-1)).
\end{eqnarray*}
Thus the lower bound is identical for any $p_j$ as it is for~$p_1$.

For $p_1 = p_j \in [1/(2^{\mu+1}-1),1/2^{\mu})$ for some $\mu$, consider
$$\left(p_1, \underbrace{\frac{1-p_1}{2^{\mu+1}-2}, \ldots,
\frac{1-p_1}{2^{\mu+1}-2}}_{2^{\mu+1}-2}\right).$$ By
Lemma~\ref{complete}, this has a complete coding tree --- in
this case with $l_{1}$ one bit shorter than the other lengths --- and
thus achieves the lower bound for this range ($\lu_j=\mu+1$).  Similarly
$$\left(p_1, \underbrace{2^{-\mu-1}, \ldots,
2^{-\mu-1}}_{2^{\mu+1}-2}, 2^{-\mu}-p_1\right)$$ has a
fixed-length optimal coding tree for $p_1 \in [1/2^\mu,
1/(2^\mu-1))$, achieving the lower bound for this range ($\lu_j=\mu$).
\end{proof}

The unit-sized bounds of (\ref{mmprbounds}) are identical to the tight
bounds at (negative integer) powers of two.  In addition, the tight
bounds clearly approach $0$ and $1$ as $p_j \downarrow 0$, similarly
to those for average redundancy\cite{MPK}.  Bounds found knowing $p_1$
are different for the two utilities, however, the average redundancy
upper and lower bounds being very close (about $0.086$
apart)\cite{Gall,Mans,MoAb}.  For larger given probabilities, note
that, above $0.5$, $p_1$ and $p_j$ bounds are identical since any such
probability must be the most probable.  Approaching $1$, the upper and
lower bounds on minimum average redundancy coding converge but never
merge, whereas the minimum maximum redundancy bounds are identical for
$p_1 \geq 2/3$.
 
\subsection{Minimized maximum pointwise redundancy codeword lengths}
\label{mmprlength}

In addition to finding redundancy bounds in terms of $p_1$ or $p_j$,
it is also often useful to find bounds on the behavior of $l_j$ in
terms of $p_j$ (for $j=1$ or general $j$), as was done for optimal
average redundancy in \cite{CaDe2} (for $j=1$).

\subsubsection*{\textbf{Lengths, Maximum Pointwise Redundancy}}
\begin{theorem}
Any code with lengths $\boldl$ minimizing $\max_{i \in \X} (l_i + \lg
p_i)$ over probability mass function~$\p$, where 
$p_j \geq 2^{-\nu}$, must have $l_j \leq \nu$.  This bound is 
tight, in the sense that,
for $p_1 < 2^{-\nu}$, one can always find a probability mass function
with $l_{1} > \nu$.  Conversely, if $p_j \leq 1/(2^{\nu}-1)$, there is
an optimal code with $l_j \geq \nu$, and this bound is also tight.
\label{mmprlen}
\end{theorem}

\begin{proof}
Suppose $p_j \geq 2^{-\nu}$ and $l_j \geq 1+\nu$.  Then
$R_\opt^*(\p) = R^*(\boldl,\p) \geq l_j + \lg p_j \geq 1$,
contradicting the unit-sized bounds of (\ref{mmprbounds}).  Thus $l_j
\leq \nu$.

For tightness of the bound, suppose $p_1 \in (2^{-\nu-1},2^{-\nu})$
and consider $n=2^{\nu+1}$ and
$$\p = \left(p_1, \underbrace{2^{-\nu-1}, \ldots, 2^{-\nu-1}}_{n-2},
2^{-\nu}-p_1\right).$$ If $l_{1} \leq \nu$, then, by the Kraft
inequality, one of $l_{2}$ through $l_{n-1}$ must exceed $\nu$.
However, this contradicts the unit-sized bounds of (\ref{mmprbounds}).
For $p_1 = 2^{-\nu-1}$, a uniform distribution results in $l_{1} =
\nu+1$.  Thus, since these two results hold for any $\nu$, this
extends to all $p_1<2^{-\nu-1}$, and this bound is tight.

Suppose $p_j \leq 1/(2^{\nu}-1)$ and consider an optimal length
distribution with $l_j < \nu$.  As in the Theorem~\ref{mmprbetter} proof, 
we consider the weights of the nodes of the corresponding extension to
the code tree at level $l_j$: $\setN$ is the set of nodes on that
level, while $\setA'$ is a set of nodes not in the tree, where each
leaf $i$ above the level has $2^{l_j-l_i}$ nodes in $\setA'$, each of
weight $p_i2^{l_i-l_j}$.  Again, the sum of the $2^{l_j}-1$ weights in 
$\setA' \cup \setN \backslash \{j\}$ is no less than $1-p_j$, so
there is one node $k'$ such that
\begin{equation}
w(k') \geq \frac{1-p_j}{2^{l_j}-1} \geq \frac{1-p_j}{2^{l_j}-2^{l_j+1-\nu}}.
\label{nuineq1}
\end{equation}
If this is in $\setN$, taking the logarithm and adding $l_j$ to 
the right-hand side,
\begin{equation}
R^*(\boldl,\p) \geq \nu-1 + \lg \frac{1-p_j}{2^{\nu-1}-1}
\label{nuineq2}
\end{equation}
the right-hand side being an upper bound to its pointwise redundancy
(based on the right-hand side of (\ref{nuineq1})).  If $k'$ is in
$\setA'$, then, using the right-hand side of (\ref{nuineq1}), the
corresponding leaf (codeword) $k$ at level $l_k < l_j$ has at least
probability
$$2^{l_j-l_k}\cdot\frac{1-p_j}{2^{l_j}-2^{l_j+1-\nu}} =
\frac{1-p_j}{2^{l_k}-2^{l_k+1-\nu}}$$ and (\ref{nuineq2}) thus still
holds.

Consider adding a bit to codeword~$j$.  Note that 
\begin{eqnarray*}
l_j+1+\lg p_j &\leq& \nu + \lg p_j \\
&\leq& \nu-1 + \lg \frac{1-p_1}{2^{\nu-1}-1}
\end{eqnarray*}
a direct consequence of $p_j \leq
1/(2^{\nu}-1)$.  Thus, if we replace this code with one for which
$l_j = \nu$, maximum redundancy is not increased and thus the new code 
is also optimal.  The tightness of the bound
is seen by applying Lemma~\ref{complete} to distributions of
the form
$$\p = \left(p_1,\underbrace{\frac{1-p_1}{2^\nu-2},
\ldots,\frac{1-p_1}{2^\nu-2}}_{2^\nu-2}\right)$$ for $p_1
\in (1/(2^{\nu}-1),1/2^{\nu-1})$.  This distribution results in $l_{1} = \nu-1$
and thus $R_\opt^*(\p) = \nu + \lg (1-p_1) - \lg (2^\nu-2)$,
which no code with $l_{1} > \nu-1$ could achieve.
\end{proof}

In particular, if $p_j \geq 0.5$, $l_j=1$, while if $p_j \leq 1/3$,
there is an optimal code with $l_j > 1$.  One might wonder about $\p =
(0.99, 0.01)$, for which two $1$-length codewords are optimal, yet $p_2
\leq 1/3$.  In this case, any code with $l_2 \leq 6$ is optimal,
having item $1$ (with $l_1 = 1$) as the item with maximum pointwise
redundancy.  Thus there is no contradiction, although this does
illustrate how this lower bound on length is not as tight as it might
at first appear, only applying to \textit{an} optimal code rather than
\textit{all} optimal codes.

\section{$d$\textsuperscript{th} Exponential Redundancy}
\label{dthbounds}

We now briefly address the $d$\textsuperscript{th} exponential
redundancy problem.  Recall that this is the minimization of (\ref{dth}),
$$R^d(\p,\boldl) = \frac{1}{d} \lg \sum_{i \in \X} p_i^{1+d} 2^{dl_{i}} =
\frac{1}{d} \lg \sum_{i \in \X} p_i 2^{d(l_{i}+\lg p_i)}.$$
A straightforward application of Lyapunov's inequality for
moments --- an application of H\"{o}lder's inequality, e.g., 
\cite[p.~27]{HLP} or \cite[p.~54]{Mitr} --- yields $R^{d'}(\p,\boldl) \leq
R^d(\p,\boldl)$ for $d' \leq d$.  Taking limits to $0$ and $\infty$,
this results in
$$
\begin{array}{ll}
0 \leq \bar{R}(\p,\boldl) \leq R^d(\p,\boldl) \leq R^*(\p,\boldl) < 1,& d > 0 \\
0 \leq R^d(\p,\boldl) \leq \bar{R}(\p,\boldl) \leq R^*(\p,\boldl) < 1,& d \in (-1,0)
\end{array}
$$
for any valid $\p$ and any $\boldl$ satisfying the Kraft inequality with
equality; the lower bound in the negative case is a result of
$$R^{-1}(\p,\boldl) = -\lg \sum_{i \in \X} 2^{-l_i} = 0,\mbox{ given }\sum_{i \in \X} 2^{-l_i} = 1.$$
This results in an extension of (\ref{mmprbounds}):
$$
\begin{array}{ll}
0 \leq \bar{R}_\opt(\p) \leq R_\opt^d(\p) \leq R_\opt^*(\p) < 1,& d > 0 \\
0 \leq R_\opt^d(\p) \leq \bar{R}_\opt(\p) \leq R_\opt^*(\p) < 1,& d \in (-1,0)
\end{array}
$$
where $R_\opt^d(\p)$ is the optimal $d$\textsuperscript{th}
exponential redundancy, an improvement on the bounds found in
\cite{Baer05}.  These inequalities lead directly to:

\subsubsection*{\textbf{Bounds, $d$\textsuperscript{th} Exponential Redundancy}}
\begin{figure}[t]
     \psfrag{R}{\Large $R_\opt^d(\p)$}
     \psfrag{p(1)}{\Large $p_1$ or $p_j$}
     \centering
          \resizebox{8.75cm}{!}{\includegraphics{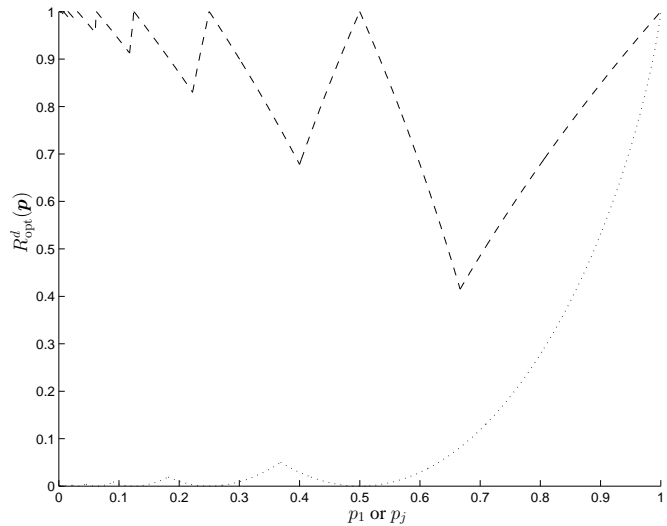}}
     \caption{Bounds on $d$\textsuperscript{th} exponential redundancy, valid for any $d>0$.
     Upper bounds dashed, lower bounds dotted.}
     \label{dar}
\end{figure}

\begin{corollary}
The upper bounds of Theorem~\ref{mmprbetter} are upper bounds for
$R_\opt^d(\p)$ with any $d$, while for $d < 0$, any upper bounds for
average redundancy (Huffman) coding will also suffice (e.g.,
\cite{Gall,Mans} for known $p_1$ or \cite{YeYe2,MPK} for any known
$p_j$).  If $d > 0$, the tight lower bounds of average redundancy
coding are lower bounds for $R_\opt^d(\p)$ with $d>0$.  These lower bounds
--- whether or not we know that $j = 1$ --- are
\begin{equation}
\bar{R}_\opt(\p) \geq \xi - (1-p_j)\lg(2^\xi-1) - H(p_j) \quad \mbox{\cite{MoAb,MPK}}
\label{moab}
\end{equation}
where $$\xi = \left\lceil \lg
\frac{1-2^\frac{1}{p_j-1}}{1-2^\frac{p_j}{p_j-1}} \right\rceil$$
for $p_j \in (0,1)$ and 
\begin{equation}
H(x) \definedas - x \lg x - (1-x) \lg (1-x).
\label{Hdef}
\end{equation}
\label{dcorollary}
\end{corollary}
This result is illustrated for $d>0$ in Fig.~\ref{dar}, showing an improvement
on the original unit bounds for values of $p_j$ other than (negative
integer) powers of two.

\section{Bounds on Exponential-Average Problems}
\label{bexp}

\subsection{Previously known exponential-average bounds}

While the average, maximum, and $d$\textsuperscript{th} average
redundancy problems yield performance bounds in terms of $p_1$ (or any
$p_j$) alone, here we seek to find any bounds on $\Cost_\q(\p,\boldl)$
in terms of $p_1$ and an appropriate entropy measure.  Such a more
general form is needed because, unlike the other objectives discussed
here, this is not a redundancy objective.

Note that $\q \leq 0.5$ is a trivial case, always solved by a finite
unary code, $$\code^{\mbox{\scriptsize u}}(n) \definedas (0, 10, 110,
\ldots, 1^{n-2}0, 1^{n-1}).$$ This can be seen by applying the
exponential combination rule (\ref{expcomb}) of the associated
Huffman-like algorithm; at each step, the combined weight will be the
lowest weight of the reduced problem, being strictly less than the
higher of the two combined weights, thus leading to a unary code.

For $\q > 0.5$, there is a relationship between this problem and R\'{e}nyi entropy.  R\'{e}nyi entropy\cite{Ren2} is defined as 
\begin{equation}
H_\alpha(\p) \definedas \frac{1}{1-\alpha}\lg \sum_{i=1}^n p_i^\alpha
\label{entropy}
\end{equation}
for $\alpha > 0$,
$\alpha \neq 1$.  It is often defined for $\alpha \in \{0,1,\infty\}$
via limits, that is, 
$$H_0(\p) \definedas \lim_{\alpha \downarrow 0} H_\alpha(\p) = \lg
\|\p\|$$ (the logarithm of the cardinality of $\p$), $$H_1(\p) \definedas
\lim_{\alpha \rightarrow 1} H_\alpha(\p) = - \sum_{i=1}^n p_i \lg
p_i$$ (the Shannon entropy of $\p$), and $$H_\infty(\p) \definedas
\lim_{\alpha \uparrow \infty} H_\alpha(\p) = -\lg p_1$$ (the
min-entropy).  

Campbell first proposed exponential utility functions for coding in
\cite{Camp0,Camp}.  He observed the simple lower bound for $\q>0.5$ in
\cite{Camp}; the simple upper bound was subsequently shown, e.g., in
\cite[p.~156]{AcDa} and \cite{BlMc}.  These bounds are similar to the
minimum average redundancy bounds.  In this case, however, the bounds
involve R\'{e}nyi's entropy, not Shannon's.

Defining
$$\alpha(\q) \definedas \frac{1}{\lg 2\q} = \frac{1}{1+\lg \q}$$
and $$\Cost_\q^\opt(\p) \definedas \min_{\boldsl \in \L} \Cost_\q(\p,\boldl)$$
the unit-sized bounds for $\q>0.5$, $\q \neq 1$ are
\begin{equation}
0 \leq \Cost_\q^\opt(\p) - H_{\alpha(\q)}(\p) < 1 .
\label{LH01}
\end{equation}
In the next subsection we show how this bound follows from a
result introduced there.

As an example of these bounds, consider the probability distribution
implied by Benford's law\cite{Newc,Benf}:
\begin{equation}
p_i = \log_{10}(i+1) - \log_{10}(i), ~ i = 1, 2, \ldots 9
\label{benf}
\end{equation}
that is,
$$\p \approx (0.30, 0.17, 0.12, 0.10, 0.08, 0.07, 0.06, 0.05, 0.05).$$
At $\q=0.6$, for example, $H_{\alpha(\q)}(\p) = 2.259\ldots$, so the
optimal code cost is between $2.259$ and~$3.260$.  In the application
given in \cite{Baer07} with (\ref{success}), these bounds correspond to an
optimal solution with probability of success (codeword transmission)
between $0.189$ and $0.316$.  Running the algorithm, the optimal
lengths are $\boldl = \ll 1,2,3,4,5,6,7,8,8 \lr$, resulting in cost
$2.382\ldots$ (probability of success $0.296\ldots$).  At $\q=2$,
$H_{\alpha(\q)}(\p) = 3.026\ldots$, so the optimal code cost is
bounded by $3.026$ and~$4.027$, while the algorithm yields an optimal
code with $\boldl = \ll 2, 3, 3, 3, 3, 4, 4, 4, 4 \lr$, resulting in
cost~$3.099\ldots$.  

The optimal cost in both cases is quite close to entropy, indicating
that better upper bounds might be possible.  In looking for better
bounds, recall first that the inequalities in (\ref{LH01}) --- like
the use of the exponential Huffman algorithm --- apply for both $\q
\in (0.5,1)$ and $\q > 1$.  Improved bounds on the optimal solution
for the $\q>1$ case are given in \cite{BlMc}, but not in closed form
or in terms of a single probability and entropy.  Closed-form
bounds for a related objective are given in \cite{Tane}.  However, the
proof for the latter set of bounds is incorrect in that it uses the
assumption that we will always have an exponential-average-optimal 
$l_1$ equal to $1$ if $p_1 \geq 0.4$.  We shortly disprove this assumption for
$\q>1$, showing the need for modified entropy bounds.  Before this, we
derive bounds based on the results of the prior section.

\subsection{Better exponential-average bounds}

Any exponential-average minimization can be transformed into a
$R^d$ minimization problem, so we can apply Corollary~\ref{dcorollary}:
Given an exponential-average
minimization problem with $\p$ and $\q$, if we define $\tilde{\alpha}
\definedas \alpha(\q) = 1/(1+\lg \q)$ and
$$\hat{p}_i \definedas \frac{p_i^{\tilde{\alpha}}}{\sum_{k=1}^n p_k^{\tilde{\alpha}}} = 
\frac{p_i^{\tilde{\alpha}}}{2^{(1-\tilde{\alpha})H_{\tilde{\alpha}}(\sp)}}$$
we have
\begin{equation}
\begin{array}{rcl}
R^{\lg \q}(\hat{\p},\boldl) 
&=& \displaystyle \frac{1}{\lg \q} \lg \sum_{i=1}^n {\hat{p}_i}^{1+\lg \q} \q^{l_i} \\
&=& \displaystyle \log_\q \sum_{i=1}^n p_i\q^{l_i} - 
\log_\q \left(\sum_{i=1}^n p_i^{\tilde{\alpha}}\right)^{\frac{1}{\tilde{\alpha}}} \\
&=& \displaystyle \Cost_\q(\boldl,\p) - H_{\tilde{\alpha}}(\p)
\end{array}
\label{trans}
\end{equation}
where $H_\alpha(\p)$ is R\'{e}nyi entropy, as in~(\ref{entropy}).
This transformation --- shown previously in \cite{BlMc} --- provides a
reduction of exponential-average minimization to
$d$\textsuperscript{th} exponential redundancy.  Thus improving bounds
for the redundancy problem improves them for the exponential-average
problem, and we can show similarly strict improvements to the
unit-sized bounds (\ref{LH01}); because $\hat{p}_j$ can be expressed
as a function of $p_j$, $\q$, and $H_{\tilde{\alpha}}(\p)$, so can
this bound:

\subsubsection*{\textbf{Bounds, Exponential-Average Objective}}
\begin{corollary}
Denote the known lower bound for optimal average redundancy
(Huffman) coding as $\bar{o}(p_j) \geq 0$ --- which is that of
Corollary~\ref{dcorollary} \cite{MoAb,MPK} ---
and the Theorem~\ref{mmprbetter} upper redundancy bound for minimized maximum
pointwise redundancy coding as $\omega^*(p_j) \leq 1$.
Further, denote the known upper
redundancy bound for optimal average redundancy given $p_1$ 
as $\bar{\omega}(p_1) \leq 1$\cite{Mans} and that given $p_j$ as 
$\bar{\omega}'(p_j) \leq 1$\cite{MPK}.  Then, for $\q > 1$, we have
\begin{eqnarray*}
\bar{o}\left(p_j^{\tilde{\alpha}} 2^{(\tilde{\alpha}-1)H_{\tilde{\alpha}}(\sp)}\right) 
&\leq& \Cost_\q^\opt(\p) - H_{\tilde{\alpha}}(\p) \\
&\leq& \omega^*\left(p_j^{\tilde{\alpha}} 2^{(\tilde{\alpha}-1)H_{\tilde{\alpha}}(\sp)}\right)
\end{eqnarray*}
Similarly, for $\q \in (0.5,1)$, we have
$$0 \leq \Cost_\q^\opt(\p) - H_{\tilde{\alpha}}(\p) \leq \bar{\omega}\left(p_1^{\tilde{\alpha}} 2^{(\tilde{\alpha}-1)H_{\tilde{\alpha}}(\sp)}\right)$$
and
$$0 \leq \Cost_\q^\opt(\p) - H_{\tilde{\alpha}}(\p) \leq \bar{\omega}'\left(p_j^{\tilde{\alpha}} 2^{(\tilde{\alpha}-1)H_{\tilde{\alpha}}(\sp)}\right).$$
\label{ecorollary}
\end{corollary}

\begin{proof}
This is a direct result of Corollary~\ref{dcorollary} and
equation~(\ref{trans}).
\end{proof}

Recall the example of Benford's distribution in (\ref{benf}) for
$\q=2$.  In this case, adding knowledge of $p_1$ improves the bounds from
$[3.026\ldots,4.026\ldots)$ to $[3.039\ldots,3.910\ldots]$ using the
$\omega^*$ from Theorem~\ref{mmprbetter} and $\bar{o}$ from \cite{MoAb}
given as (\ref{moab}) here.  For $\q=0.6$, the bounds on cost are reduced
from $[2.259\ldots,3.259\ldots)$ to $[2.259\ldots,2.783\ldots]$
using $\bar{\omega}$ given as (10) in \cite{Gall}:
$$\bar{R}_\opt(\tilde{\p}) \leq 2 - H(\tilde{p}_1) - \tilde{p}_1$$ with argument
$$\tilde{p}_1 = p_1^{\tilde{\alpha}} 2^{(\tilde{\alpha}-1)H_{\tilde{\alpha}}(\sp)} = 0.8386\ldots.$$
Recall from (\ref{Hdef}) that $$H(x) = - x \lg x - (1-x) \lg (1-x).$$

Although the bounds derived from Huffman coding are close for $\q
\approx 1$ (the most common case), these are likely not tight bounds;
we introduce another bound for $\q < 1$ after deriving a certain
condition in the next section.

\subsection{Exponential-average shortest codeword length}

Techniques for finding Huffman coding bounds do not always translate
readily to exponential generalizations because R\'{e}nyi entropy's
very definition\cite{Ren2} involves a relaxation of a property used in
finding bounds such as Gallager's entropy bounds\cite{Gall}, namely
\begin{eqnarray*}
\lefteqn{H_1[t p_1, (1-t)p_1, p_2, \ldots, p_n] =} \\ 
&& H_1[p_1, p_2, \ldots, p_n] + p_1 H_1(t,1-t)
\end{eqnarray*}
for Shannon entropy $H_1$ and $t \in [0,1]$.  This fails to hold
for R\'{e}nyi entropy.  The penalty function $\Cost_\q$ differs from
the usual measure of expectation in an analogous fashion, and we
cannot know the weight of a given subtree in the optimal code (merged
item in the coding procedure) simply by knowing the sum probability of
the items included.  However, we can improve upon the 
Corollary~\ref{ecorollary} bounds for the exponential problem 
when we know that $l_{1}=1$; the question then
becomes when we can know this given only $\q$ and $p_1$:

\subsubsection*{\textbf{Length $l_1 = 1$, Exponential-Average Objective}}
\begin{theorem}
There exists a code minimizing $\Cost_\q(\p,\boldl) \definedas
\log_\q \sum_{i \in \X} p_i \q^{l_{i}}$ with $l_{1}=1$ for $\q$ and $\p$ if
either $\q \leq 0.5$ or both $\q \in (0.5,1]$ and $p_1 \geq
2\q/(2\q+3)$.  Conversely, given $\q \in (0.5,1]$ and $p_1 <
2\q/(2\q+3)$, there exists a $\p$ such that any code with $l_{1}=1$ is
suboptimal.  Likewise, given $\q > 1$ and $p_1 < 1$, there
exists a $\p$ such that any code with $l_{1}=1$ is suboptimal.  
\label{l1}
\end{theorem}

\begin{figure}[t]
     \centering
               \psfrag{theta}{\mbox{\huge $\q$}}
               \psfrag{p1}{\mbox{\huge $p_1$}}
               \psfrag{0.0}{\mbox{\Large $0.0$}}
               \psfrag{0.2}{\mbox{\Large $0.2$}}
               \psfrag{0.4}{\mbox{\Large $0.4$}}
               \psfrag{0.6}{\mbox{\Large $0.6$}}
               \psfrag{0.8}{\mbox{\Large $0.8$}}
               \psfrag{1.0}{\mbox{\Large $1.0$}}
          \resizebox{7cm}{!}{\includegraphics{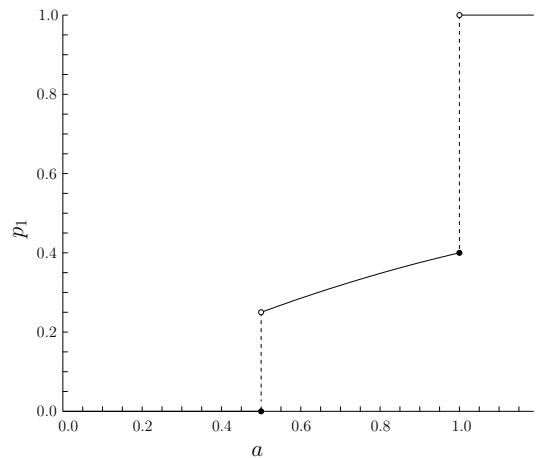}}
     \caption{Minimum $p_1$ sufficient for the existence of an optimal $l_{1}$ not exceeding $1$.}
     \label{p1}
\end{figure}

\begin{proof}
Recall that the exponential Huffman algorithm combines the items with
the smallest weights, $w'$ and~$w''$, yielding a new item of weight $w
= \q w'+\q w''$, and this process is repeated on the new set of weights,
the tree thus being constructed up from the leaves to the root.  This
process makes it clear that, as mentioned, the finite unary code (with
$l_{1}=1$) is optimal for all $\q \leq 0.5$.  This leaves the two
nontrivial cases.

\subsubsection{$\q \in (0.5,1]$}
The proof in this case is a generalization of \cite{John} and is only
slightly more complex to prove.  Consider the coding step at which
item $1$ gets combined with other items; we wish to prove that this is
the last step.  At the beginning of this step the (possibly merged)
items left to combine are $\{1\}, S_2^k, S_3^k, \ldots, S_k^k$, where
we use $S_j^k$ to denote both a (possibly merged) item of weight
$w(S_j^k)$ and the set of (individual) items combined in item~$S_j^k$.
Since $\{1\}$ is combined in this step, all but one $S_j^k$ has at
least weight $p_1$ (reflected in the second inequality below).  Note
too that all weights $w(S_j^k)$ must be less than or equal to the sums
of probabilities $\sum_{i \in S_j^k} p_i$ (reflected in the third
inequality below); equality only occurs for when $S_j^k$ has a single
item, due to multiplication by $\q < 1$ upon each merge step.  Then
$$
\begin{array}{rclll}
\frac{2\q(k-1)}{2\q+3} &\leq& (k-1) p_1 && \\
&<& p_1 + \sum_{j=2}^k w(S_j^k) && \\
&\leq& p_1 + \sum_{j=2}^k \sum_{i \in S_j^k} p_i && \\
&=& \sum_{i=1}^n p_i &=& 1
\end{array}
$$ which, since $\q > 0.5$, means that $k < 5$.  Thus we can ignore
all merging steps prior to having four items and begin with this step;
if we start out with fewer than four items ($n \leq 3$), we are
guaranteed an optimal code with $l_{1}=1$.  Four items remain, one
of which is item $\{1\}$ and the others of which are $S_2^4$, $S_3^4$,
and~$S_4^4$.  We show that, if $p_1 \geq 2\q/(2\q+3)$, these items are
combined as shown in Fig.~\ref{unary}.

\begin{figure}[t]
     \centering
               \psfrag{X}{\mbox{\small $\X$}}
               \psfrag{p1}{\mbox{\small $\{1\}$}}
               \psfrag{S22}{\mbox{\small $S_2^2$}}
               \psfrag{S2}{\mbox{\small $S_2^4$}}
               \psfrag{W2}{\mbox{\scriptsize $\begin{array}{l} \mbox{If } |S_2^4|>1, \\ \quad w(S_2^4) = \\ \qquad \q w(S'_2)+\q w(S''_2)\end{array}$}}
               \psfrag{S2'}{\mbox{\small $S'_2$}}
               \psfrag{S2''}{\mbox{\small $S''_2$}}
               \psfrag{S3}{\mbox{\small $S_3^4$}}
               \psfrag{S4}{\mbox{\small $S_4^4$}}
               \psfrag{S3,4}{\mbox{\small \thinspace $S_3^4 \cup S_4^4$}}
               \psfrag{W3,4}{\mbox{\scriptsize $\begin{array}{l} w(S_3^4 \cup S_4^4) = \\ \quad \q w(S_3^4)+\q w(S_4^4) \end{array}$}}
          \resizebox{5.5cm}{!}{\includegraphics{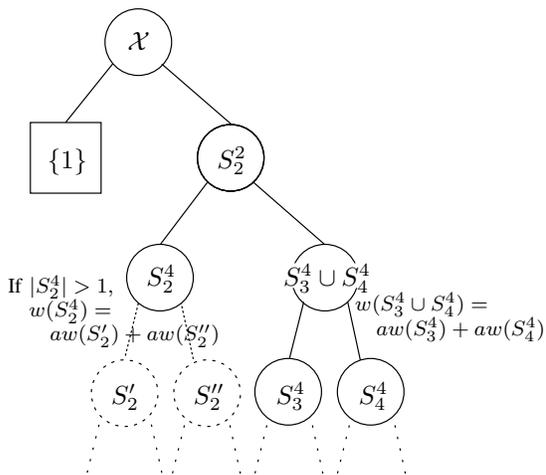}}
     \caption{Tree in last steps of the exponential Huffman algorithm.}
     \label{unary}
\end{figure}

We assume without loss of generality that weights $w(S_2^4)$,
$w(S_3^4)$, and $w(S_4^4)$ are in descending order.  From 
\begin{eqnarray*}
w(S_2^4) + w(S_3^4) + w(S_4^4) &\leq& \sum_{i=2}^n p_i \\
&\leq& \frac{3}{2\q+3} \text{,} \\
w(S_2^4) &\geq& w(S_3^4) \text{,} \\
\text{and } w(S_2^4) &\geq& w(S_4^4)
\end{eqnarray*}
it follows that $w(S_3^4) + w(S_4^4) \leq 2/(2\q+3)$.  Consider set~$S_2^4$.  
If its cardinality is $1$, then
\begin{equation}
p_1 \geq w(S_2^4) \geq w(S_3^4) \geq w(S_4^4)
\label{nec1}
\end{equation}
so the next step merges the least two weighted items $S_3^4$ and~$S_4^4$.
Since the merged item has weight at most $2\q/(2\q+3)$, this item can
then be combined with $S_2^4$, then $\{1\}$, so that $l_1=1$.  If
$S_2^4$ is a merged item, let us call the two items (sets) that merged
to form it $S'_2$ and $S''_2$, indicated by the dashed nodes in
Fig.~\ref{unary}.  Because these were combined prior to this step,
$$w(S'_2)+w(S''_2) \leq w(S_3^4)+w(S_4^4)$$ so 
$$w(S_2^4) \leq \q w(S_3^4)+\q w(S_4^4) \leq \frac{2\q}{2\q+3}.$$
Thus (\ref{nec1}) still applies, and, as in the other case, $l_1=1$.

This can be shown to be tight by noting that, for any $\epsilon \in
(0, (2\q-1)/(8\q+12))$,
$$\textstyle \p^{(\epsilon)} \definedas \left(\frac{2\q}{2\q+3}-3\epsilon,
\frac{1}{2\q+3}+\epsilon, \frac{1}{2\q+3}+\epsilon,
\frac{1}{2\q+3}+\epsilon\right)$$ achieves optimality only with length
vector $\boldl = \ll 2,2,2,2 \lr$.  The result extends to smaller~$p_1$.

\subsubsection{$\q > 1$}
Given $\q>1$ and $p_1 < 1$, we
wish to show that a probability distribution always exists such that
there is no optimal code with $l_{1} = 1$.  We first show that, for the
exponential penalties as for the traditional Huffman penalty, every
optimal $\boldl$ can be obtained via the (modified) Huffman procedure.
That is, if multiple length vectors are optimal, each optimal length
vector can be obtained by the Huffman procedure as long as ties are
broken in a certain manner.

Clearly the optimal code is obtained for $n=2$.  Let $n'$ be the
smallest $n$ for which there is an $\boldl$ that is optimal but cannot
be obtained via the algorithm.  Since $\boldl$ is optimal, consider
the two smallest probabilities, $p_{n'}$ and $p_{n'-1}$.  In this
optimal code, two items having these probabilities (although not
necessarily items $n'-1$ and $n'$) must have the longest codewords and
must have the same codeword lengths.  Were the latter not the case,
the codeword of the more probable item could be exchanged with one of
a less probable item, resulting in a better code.  Were the former not
the case, the longest codeword length could be decreased by one
without violating the Kraft inequality, resulting in a better code.
Either way, the code would no longer be optimal.  Thus we can
find two such smallest items with largest codewords (by breaking any
ties properly), which, without loss of generality, can be considered
siblings.  Therefore the problem can be reduced to one of size
$n'-1$ via the exponential Huffman algorithm.  But since all problems of
size $n'-1$ can be solved via the algorithm, this is a contradiction,
and the Huffman algorithm can thus find any optimal code.

Note that this is not true for minimizing maximum pointwise
redundancy, as the exchange argument no longer holds.  This is why the
sufficient condition of Section~\ref{bred} was not verified using
Huffman-like methods.

Now we can show that there is always a code with $l_{1}>1$ for any
$p_1 \in (0.2,1)$; $p_1 \leq 0.2$ follows easily.  Let
$$m = \left\lfloor \log_{\q} \left(\frac{4p_1}{1-p_1}\right)
\right\rfloor$$ and suppose $n = 1 + 2^{2+m}$ and $p_i =
(1-p_1)/(n-1)$ for all $i \in \{2, 3, \ldots, n\}$.  Although item $1$
need not be merged before the penultimate step, at this step its weight
is strictly less than either of the two other remaining weights, which 
have values $w' = \q^{1+m}(1-p_1)/2$.  This distribution has an optimal 
code only with $l_{1} \geq 2$.  (This must be an equality unless $m$ is 
equal to the logarithm from which it is derived, in which case $l_1$
can be either $2$ or~$3$.)  Thus, knowing merely the values of $\q>1$ and
$p_1<1$ is not sufficient to ensure that $l_{1}=1$.
\end{proof}

These relations are illustrated in Fig.~\ref{p1}, a plot of the
minimum value of $p_1$ sufficient for the existence of an optimal
code $\boldl^\opt$ with $l_1^\opt$ not exceeding~$1$.

Similarly to minimum maximum pointwise redundancy, we can observe
that, for $\q \geq 1$ (that is, $\q > 1$ and traditional Huffman
coding), a necessary condition for $l_1^\opt=1$ is $p_1 \geq 1/3$.  The
sum of the last three combined weights is at least~$1$, and $p_1$
must be no less than the other two.  However, for $\q < 1$, there is
no such necessary condition for~$p_1$.  Given $\q \in (0.5,1)$ and $p_1 \in
(0,1)$, consider the probability distribution consisting of one item
with probability $p_1$ and $n=1+2^{1+g}$ items with equal
probability, where
$$g = \max\left(\left\lfloor \log_a \frac{2ap_1}{1-p_1}
\right\rfloor, \left\lfloor \lg \frac{1-2p_1}{p_1} \right\rfloor ,0
\right)$$ and, by convention, we define the logarithm of negative
numbers to be~$-\infty$.  Setting $p_i = (1-p_1)/(n-1)$ for all $i
\in \{2, 3, \ldots, n\}$ results in a monotonic probability mass
function in which $(1-p_1)a^g/2 < p_1$, which means that the
generalized Huffman algorithm will have in its penultimate step three
items: One of weight $p_1$ and two of weight $(1-p_1)a^g/2$; these
two will be complete subtrees with each leaf at depth~$g$.  Since
$(1-p_1)a^g/2 < p_1$, $l_1^\opt=1$.  Again, this holds for any $\q \in
(0.5,1)$ and $p_1 \in (0,1)$, so no nontrivial necessary condition
exists for $l_1^\opt=1$.  It is also the case for $\q \leq 0.5$, since
the unary code is optimal for any probability mass function.

\subsection{Exponential-average bounds for $\q \in (0.5,1)$, $p_1 \geq 2\q/(2\q+3)$}

Entropy bounds derived from Theorem~\ref{l1}, although rather
complicated, are, in a certain sense, tight:

\subsubsection*{\textbf{Further Bounds, Exponential-Average Objective}}
\begin{corollary}
In the minimization of $\Cost_\q(\p,\boldl) \definedas
\log_\q \sum_{i \in \X} p_i \q^{l_{i}}$, if $\q \in (0.5,1)$ and a 
minimizing $\boldl$ has $l_1 = 1$ (i.e., all $p_1 \geq 2\q/(2\q+3)$),
the following inequalities hold, where $\tilde{\alpha} = \alpha(\q) \definedas 
1/(1+\lg \q)$:
$$\sum_{i=1}^n p_i \q^{l_{i}} > \q^2 \left[ 
\q^{{\tilde{\alpha}} H_{\tilde{\alpha}}(\sp)} - p_1^{\tilde{\alpha}} \right]^{\frac{1}{\tilde{\alpha}}} + \q p_1$$
or, equivalently,
$$\Cost_\q(\p) < 1 + \log_\q \left(\q \left[ \q^{{\tilde{\alpha}} H_{\tilde{\alpha}}(\sp)} - p_1^{\tilde{\alpha}} \right]^{\frac{1}{\tilde{\alpha}}} + p_1\right)$$
and
$$\sum_{i=1}^n p_i \q^{l_{i}} \leq \q \left[\q^{{\tilde{\alpha}} H_{\tilde{\alpha}}(\sp)} - p_1^{\tilde{\alpha}} \right]^{\frac{1}{\tilde{\alpha}}} + \q p_1$$
or, equivalently,
$$\Cost_\q(\p) \geq 1 + \log_\q \left(\left[ \q^{{\tilde{\alpha}} H_{\tilde{\alpha}}(\sp)} - p_1^{\tilde{\alpha}} \right]^{\frac{1}{\tilde{\alpha}}} + p_1\right).$$
\label{coro}
\end{corollary}
This upper bound is tight for $p_1 \geq 0.5$ in the sense
that, given values for $\q$ and $p_1$, we can find $\p$ to make the inequality 
arbitrarily close.  Probability distribution $\p = (p_1, 1-p_1 +\epsilon,
\epsilon)$ does this for small~$\epsilon$, while the lower bound is
tight (in the same sense) over its full range, since $\p = (p_1,
(1-p_1)/4, (1-p_1)/4, (1-p_1)/4, (1-p_1)/4)$ achieves it (with a
zero-redundancy subtree of the weights excluding~$p_1$).

\begin{proof}
We apply the simple unit-sized coding bounds (\ref{LH01}) for the subtree
that includes all items but item $\{1\}$.  Let $B = \{2, 3, \ldots, n\}$
with $p_i^B = \P[i~|~i \in B] = p_i/(1-p_1)$ and with R\'{e}nyi $\alpha$-entropy
$$H_{\tilde{\alpha}}(\p^B) = \frac{1}{1-{\tilde{\alpha}}} \lg \sum_{i=2}^n
{\left(\frac{p_i}{1-p_1}\right)}^{\tilde{\alpha}}.$$
$H_{\tilde{\alpha}}(\p^B)$ is related to the entropy of the original
source $\p$ by
$$2^{(1-{\tilde{\alpha}})H_{\tilde{\alpha}}(\sp)} = (1-p_1)^{\tilde{\alpha}}
2^{(1-{\tilde{\alpha}})H_{\tilde{\alpha}}(\sp^B)} + p_1^{\tilde{\alpha}}$$
or, equivalently, since $2^{1-{\tilde{\alpha}}} = \q^{\tilde{\alpha}}$, 
\begin{equation}
\q^{H_{\tilde{\alpha}}(\sp^B)} = \frac{1}{1-p_1} \left[\q^{{\tilde{\alpha}} H_{\tilde{\alpha}}(\sp)}
  - p_1^{\tilde{\alpha}}\right]^\frac{1}{\tilde{\alpha}}.
\label{entropyB}
\end{equation}
Applying (\ref{LH01})
to subtree $B$, we have 
$$\q^{H_{\tilde{\alpha}}(\sp^B)} \geq \frac{1}{(1-p_1)\q}\sum_{i=2}^n p_i\q^{l_{i}}
> \q^{H_{\tilde{\alpha}}(\sp^B)+1}.$$  The bounds for $\sum_i p_i\q^{l_{i}}$ are
obtained by substituting (\ref{entropyB}), multiplying both sides by $(1-p_1)\q$, and adding the contribution of item $\{1\}$, $\q p_1$.
\end{proof}

A Benford distribution (\ref{benf}) for $\q=0.6$ yields
$H_{\alpha(\q)}(\p) \approx 2.260$.  Since $p_1 > 2\q/(2\q+3)$,
$l_{1}$ is~$1$ and the probability of success is between $0.250$ and
$0.298$; that is, $\Cost_\q^\opt \in [2.372\ldots,2.707\ldots)$.
Recall that the bounds found using (\ref{trans}) were
$\P[\mbox{success}] \in (0.241,0.316)$ and $\Cost_\q^\opt \in
[2.259\ldots, 2.783\ldots]$, an improvement on the unit-sized
bounds, but not as good as those of Corollary~\ref{coro}.  The optimal code
$\boldl = \ll 1,2,3,4,5,6,7,8,8 \lr$ yields a probability of success
of $0.296$ ($\Cost_\q^\opt=2.382\ldots$).

Note that these arguments fail for $\q > 1$ due to the lack of
sufficient conditions for $l_{1} = 1$.  For $\q < 1$, other cases
likely have improved bounds that could be found by bounding $l_{1}$
--- as with the use of lengths in \cite{MoKu} to come up with general
bounds\cite{CaDe1, Mans} --- but new bounds would each cover a more
limited range of $p_1$ and be more complicated to state and to prove.

\section*{Acknowledgment}

The author would like to thank J. David Morgenthaler for discussions on this topic.

\ifx \cyr \undefined \let \cyr = \relax \fi

\end{document}